\documentstyle[psfig,conf_iap,]{article}
\begin{document}
\heading{%
Uncloaking the High Redshift Galaxy Population With The
Complete Optical and Radio Absorption Line System (CORALS) Survey.
} 
\par\medskip\noindent
\author{%
Sara L. Ellison$^{1}$, Lin Yan$^{2}$, Isobel M. Hook$^{3}$, Max Pettini$^{4}$,
Jasper V. Wall$^{3}$, Peter Shaver$^{5}$
}
\address{European Southern Observatory, Santiago, CHILE}
\address{SIRTF Science Center, Caltech, California, USA}
\address{Oxford University, Oxford, UK}
\address{Institute of Astronomy, Cambridge, UK}
\address{European Southern Observatory, Garching bei Munchen, GERMANY}

\begin{abstract}
We present the first results from the Complete Optical and Radio 
Absorption Line System (CORALS) survey for Damped
Lyman Alpha (DLA) systems based on a radio-selected sample of
quasars.  We find that the number density and
neutral gas content of CORALS DLAs is in good agreement with
previous surveys.  These results indicate that
dust obscuration has probably not been responsible for a significant
selection bias in the past, although the statistics still permit
an underestimate of $n(z)$ and $\Omega_{DLA}$ by up to a factor
of about 2.
\end{abstract}

\section{Introduction: The CORALS Survey}
It is widely recognised that DLAs represent one of the most promising
methods of probing the high redshift galaxy population.  However,
it is similarly acknowledged that the presence of dust in these
galaxies could impose a significant bias on our current database of
DLAs, which are mostly drawn from optically bright QSO samples.
Observational evidence for such a concern comes, for example, from 
statistically steeper continuum slopes in QSOs with DLAs \cite{pfb91} 
and lack of metallicity evolution in DLAs
from $0.5 < z_{abs} < 3.5$ \cite{pet99}.

The CORALS survey
aims to address this concern by using a radio-selected sample of QSOs
to search for DLAs.  Full details of this sample can be found in 
\cite{corals}.  In brief, a complete ($S_{2.7 GHz} > 0.25$ Jy)
sample of $z_{em} \geq 2.2$ QSOs was culled from the Parkes Catalogue.
Optical counterparts have been identified for every target, resulting in
a sample of 66 radio-selected QSOs with no optical magnitude limit.
Follow-up spectroscopy revealed 22 $1.8 < z_{abs} < z_{em}$
DLAs, of which 3 are excluded from our statistics
due to their proximity to the QSO 
($\Delta v <$ 3000 km/s).

\section{Results}
In order to assess the extent to which previous surveys may have
been biased due to dust, we calculate the number density, $n(z)$, and
neutral gas mass density, $\Omega_{DLA}$, of CORALS DLAs.  
We find that $\log \Omega_{\rm DLA} h_{65} =  -2.59^{+0.17}_{-0.24}$ 
($\Omega_M = 1.0, \Omega_{\Lambda} = 0$) over the redshift range
$1.8 < z_{abs} < 3.5$, in
good agreement with the latest results from \cite{per},
see Figure 1.  Perhaps most importantly, we do not find any evidence
for a previously undetected population of high column density
absorbers.  Similarly, the number density of CORALS DLAs $n(z) =
0.31^{+0.09}_{-0.08}$ at a mean absorption redshift $\langle
z_{abs} \rangle= 2.37$ is 50\% larger than previous surveys (e.g.
\cite{lsl00}).  However, the difference is 
significant only at around the 1$\sigma$ level.  Overall, we conclude
that previous surveys may have underestimated $n(z)$ and $\Omega_{DLA}$
by at most a factor of two.

\begin{figure}
\centerline{\vbox{
\psfig{figure=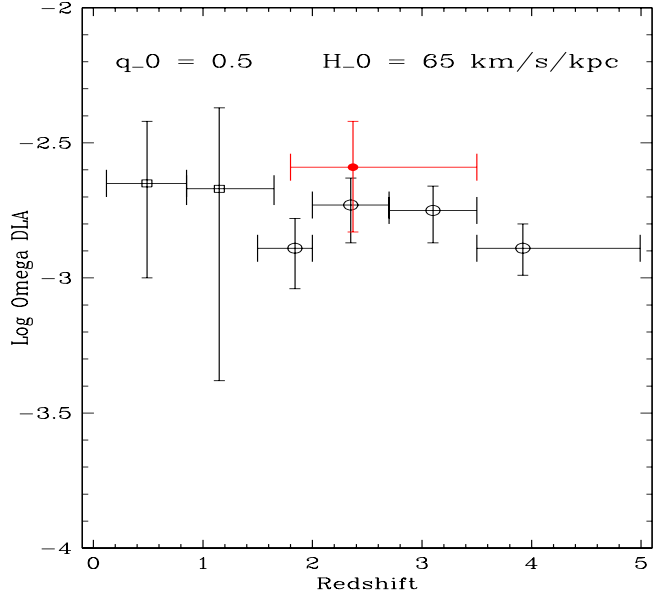,height=8.5cm,width=9cm}
}}
\caption[]{$\Omega_{DLA}$ measurements taken from: \cite{per} (open
circles), \cite{rt00} (open squares) and CORALS \cite{corals} (solid
circle).}
\end{figure}
%



%
\vfill
\end{document}